\documentclass[12pt,reqno]{article}

\usepackage[american]{babel}
\usepackage{csquotes}
\usepackage[style=apa, 
backend=biber,
sortcites=true,
sorting=nyt]{biblatex}

\DeclareLanguageMapping{american}{american-apa}
\usepackage{amsmath, amssymb, amsfonts, amsthm}

\usepackage{graphicx}
\usepackage{epsfig}
\usepackage{graphics}
\usepackage{epstopdf}
\usepackage{caption}
\usepackage{subcaption}
\usepackage{tikz}
\usetikzlibrary{positioning,arrows.meta,shapes}

\usepackage{tabularx}
\usepackage{array}
\usepackage{booktabs}
\usepackage{makecell}
\usepackage{multirow}
\usepackage{threeparttable}
\usepackage{siunitx}
\usepackage[margin=1in]{geometry}
\usepackage{setspace}

\usepackage{algorithm}
\usepackage{algorithmicx}
\usepackage{algpseudocode}

\usepackage{enumitem}      
\usepackage{float}         
\usepackage{url}           
\usepackage{hyperref}      

\hypersetup{
	colorlinks=true,
	linkcolor=blue,
	citecolor=blue,
	urlcolor=blue
}

\usepackage{authblk}

\title{Starshaped Mean Residual Life Models for Non-Monotonic Survival Data: 
	A Bayesian PMRL Regression Framework with Applications to Teacher Retention}

\author[1]{Mohammad Sepehrifar\thanks{msepehrifar@math.msstate.edu}}

\affil[1]{Department of Mathematics and Statistics, Mississippi State University, 175 President Circle, Mississippi State, MS 39762, USA}

\date{\today}

\begin{document}
	\maketitle

\begin{abstract}
	We develop a Starshaped Mean Residual Life (SMEL) framework for survival data 
	with non-monotonic hazard patterns, where early-stage attrition is followed by 
	mid-career stabilization. Unlike Cox proportional hazards models or standard mean 
	residual life models requiring monotonicity, SMEL accommodates complex temporal 
	dynamics by requiring only that $m(t)/t$ be nondecreasing, formalizing the 
	transition from vulnerability to equilibrium. We extend SMEL to regression 
	settings via proportional mean residual life (PMRL) models, 
	$m(t\mid Z)=m_0(t)\exp(Z^\top\gamma)$, with adaptive Bayesian estimation using 
	three-parameter Weibull--resilience distributions and the No-U-Turn Sampler. 
	Monte Carlo simulations across 48,000 datasets show SMEL-PMRL maintains bias 
	$\leq 0.02$ under 40\% right-censoring, reduces integrated Brier score by 19\% 
	over Cox models ($2.34$ vs.\ $2.88\times10^{-2}$), and achieves 5.4\% AIC 
	improvement. Joint longitudinal-survival extensions via shared frailty enable 
	simultaneous modeling of correlated time-to-event and continuous outcomes. 
	Application to 169 rural STEM teachers (2018--2023, NSF Noyce) confirms 
	starshaped equilibrium ($\Lambda=12.47$, $p=0.002$), with 38\% early-career 
	tenure decline (years 1--3). The joint model ($\hat{\theta}=0.41$, 95\% CI: 
	$[0.35,\,0.47]$) shows persistence beyond year~3 yields 31-point cumulative 
	achievement gains (0.56~SD) over four years. SMEL-PMRL offers a flexible, 
	theoretically grounded alternative to proportional hazards for workforce dynamics 
	and high-attrition settings where equilibrium processes govern long-term 
	stability.
\end{abstract}


		

\footnotetext{Correspondence: msepehrifar@math.msstate.edu
	}



%
%
%
%

\section{Introduction}
		Persistent shortages of qualified STEM educators in rural school districts represent one of the most formidable barriers to achieving educational equity in the United States. National data indicate that rural schools experience teacher turnover rates up to 50\% higher than urban districts within the first five years of service \parencite{Ingersoll2012, Sutcher2019}, with the disparity particularly pronounced in science, technology, engineering, and mathematics subjects where alternative career opportunities in private industry exert competitive pressure on the teaching workforce. Geographic isolation, limited access to professional development, restricted mentorship infrastructure, and systemic challenges in teacher recruitment collectively contribute to enduring disparities in instructional quality and student learning outcomes in rural contexts \cite{Marder2022, Showalter2019}.
		
		Despite considerable federal investment through initiatives such as the National Science Foundation's Robert Noyce Teacher Scholarship Program---which has supported over 10,000 STEM teachers since 2002---the field continues to lack robust quantitative frameworks capable of predicting and mitigating the nonlinear attrition patterns characteristic of rural teaching careers. Traditional survival analysis approaches, while informative, impose restrictive assumptions regarding proportional hazards or parametric distributional forms that frequently fail to capture the complex temporal dynamics inherent in teacher workforce systems \cite{Borman2008}. Recent methodological advances in educational data mining and learning analytics have expanded capacity to model intricate relationships \cite{Koedinger2015}, yet these approaches often prioritize predictive accuracy over theoretical interpretability and may inadequately address the temporal structure central to career trajectory analysis.
		This study introduces the Starshaped Mean Residual Life (SMEL) framework---originally developed in reliability engineering to model system aging and component failure processes \cite{Sepehrifar2025}---to educational workforce research. We adapt SMEL to analyze teacher retention and student achievement as coupled dynamic equilibrium processes, where ``equilibrium'' denotes a state in which expected remaining career duration stabilizes after an initial high-vulnerability period. The framework's theoretical foundation rests on a formal property: the ratio of mean residual life to elapsed time must be nondecreasing, ensuring that relative remaining career expectancy improves or stabilizes following early-career ``burn-in'' attrition. This starshaped property provides a mathematically rigorous criterion for identifying when workforce systems transition from precarious disequilibrium to sustainable retention equilibrium---a transition obscured by conventional survival models.

\subsection{Teacher Retention in High-Need Contexts}

%

 Early-career attrition represents a critical vulnerability in the teacher workforce, with approximately 44\% of beginning teachers exiting the profession within five years \cite{Ingersoll2001, Ingersoll2012}. Teacher turnover generates substantial financial costs—estimated at $10{,}000–$20{,}000 per replacement—and disrupts instructional continuity, erodes institutional knowledge, and disproportionately harms schools serving disadvantaged students \cite{CarverThomas2017, Ronfeldt2013}.
 Rural schools face compounded retention challenges. Geographic isolation restricts access to professional networks and spousal employment opportunities \cite{Podolsky2019}, while teachers frequently assume multiple roles that intensify workload and elevate burnout risk. Compensation differentials further disadvantage rural districts competing with urban schools and private-sector employers \cite{Goldhaber2011, Showalter2019}. Retention decisions emerge from complex interactions between individual characteristics and institutional conditions, including administrative support and resource adequacy \cite{CarverThomas2017, Nguyen2024}. However, most quantitative studies rely on linear or Cox proportional hazards models that impose constant covariate effects over time—an assumption often violated by the pronounced peak in attrition risk during early career stages \cite{Borman2008}.

\subsection{Student Achievement Consequences of Teacher Turnover}

Teacher stability affects student learning through multiple pathways. Instructional continuity allows for cumulative skill development and refined pedagogy, while sustained teacher--student relationships facilitate differentiated instruction and strengthen academic trust, which is particularly important in STEM disciplines \cite{DarlingHammond2010, Ronfeldt2013}.

Empirical evidence consistently documents achievement penalties associated with turnover. A one standard deviation increase in school-level turnover reduces student achievement by approximately 0.10 standard deviations, with effects concentrated in schools serving low-income and minority students \cite{Ronfeldt2013}. Teacher experience, especially in the first five years, predicts learning gains, with the steepest growth occurring between years one and three \cite{Harris2011, Papay2015}.

Rural contexts amplify these consequences. In small schools, a single departure can eliminate essential collaborative planning, and limited labor markets often force the hiring of underqualified replacements, creating negative feedback loops that degrade school climate \cite{Kraft2016}. Despite this evidence, research has not systematically examined how the \emph{timing and trajectory} of attrition risk interacts with student achievement over time. Understanding whether effects stem from continuity or from the selective retention of more effective teachers is crucial for intervention design. The SMEL framework's joint modeling capability allows for empirical differentiation between these mechanisms.

\subsection{Limitations of Existing Quantitative Approaches}

Conventional statistical models for teacher retention rely on assumptions that poorly match the complex temporal dynamics of educational careers. Three limitations motivate the proposed methodological advance.

\paragraph{Proportional Hazards Violations.}
Cox proportional hazards models, widely used in retention research \cite{Goldhaber2011, Nguyen2024}, assume time-invariant hazard ratios:
\[
\frac{h(t \mid \mathbf{Z}_1)}{h(t \mid \mathbf{Z}_2)}
= \exp\!\left[(\mathbf{Z}_1 - \mathbf{Z}_2)^\top \boldsymbol{\beta}\right].
\]
Educational data frequently violate this assumption, as early-career teachers face elevated attrition risk that declines among persisters \cite{Borman2008}. Schoenfeld residual diagnostics often detect these violations, rendering standard Cox inference biased and unreliable. Extensions with time-varying coefficients reduce interpretability, while parametric alternatives (e.g., Weibull, log-logistic) impose distributional constraints that can miss the non-monotonic hazard patterns characteristic of teacher careers.

\paragraph{Limited Policy Relevance of Hazard Metrics.}
Traditional survival models emphasize instantaneous exit risk rather than expected remaining career duration. The mean residual life (MRL) function directly addresses policy-relevant questions: ``Given persistence to year $t$, how many additional service years can we expect?'' Formally,
\[
m(t) = E[T - t \mid T > t]
= \frac{\int_t^\infty S(u)\,du}{S(t)},
\]
where $T$ denotes career duration and $S(t)$ the survival function. MRL supports hiring, budgeting, and cost-effectiveness analyses without the cognitive translation required for hazard ratios. Although reliability theory \cite{Hall1981, Guess1986} and proportional MRL regression \cite{Maguluri1994} provide a strong foundation, educational applications remain scarce. Existing MRL models typically impose monotonicity or require pre-specified turning points (e.g., DIMRL), limiting their ability to capture nonlinear resilience. The SMEL framework relaxes these constraints by requiring only that $m(t)/t$ be nondecreasing, formalizing the transition from early-career disequilibrium to sustainable equilibrium without arbitrary thresholds.

\paragraph{Absence of Joint Teacher--Student Modeling.}
Educational production functions treat teacher labor as a fixed input \cite{Hanushek2010}, while retention studies rarely model student achievement as both a determinant and consequence of persistence \cite{CarverThomas2017}. This separation obscures feedback mechanisms: teacher effectiveness influences retention decisions, and stability affects achievement through instructional continuity. Joint longitudinal--survival models with shared random effects \cite{Hsieh2006, Rizopoulos2012}, standard in medical research, remain underused in education despite their ability to quantify these coupled dynamics within hierarchical structures.

\subsection{The Starshaped Mean Residual Life (SMEL) Framework}

We address the limitations articulated above by adapting the SMEL framework from reliability engineering to educational workforce dynamics. The approach reconceptualizes teacher retention not as a hazard-driven failure process but as an evolving equilibrium between investment forces (human capital accumulation, community integration, professional identity development) and exit pressures (workload demands, geographic isolation, opportunity costs).

\subsubsection{Starshaped Property as Equilibrium Criterion}

A lifetime distribution belongs to the SMEL class if its mean residual life function satisfies the \emph{starshaped property}: the ratio $m(t)/t$ is nondecreasing in $t$. Formally, for all $0 \leq t_1 < t_2$,
\[
\frac{m(t_1)}{t_1} \leq \frac{m(t_2)}{t_2}.
\]
This condition captures the transition from early-career vulnerability—where expected remaining tenure $m(t)$ may decline rapidly—to sustainable equilibrium where the ratio $m(t)/t$ stabilizes or increases. Unlike monotonic MRL classes (DMRL, IMRL) or models requiring pre-specified turning points (DIMRL) \cite{Guess1986}, SMEL accommodates initial steep declines followed by gradual stabilization without arbitrary thresholds \cite{Sepehrifar2025}. This flexibility is essential for modeling teacher careers exhibiting burn-in attrition (years 1--3) followed by mid-career stability.

\paragraph{\textbf{Educational Interpretation.}}
The starshaped property formalizes three career phases: (I) \emph{Burn-in}, where reality shock and competing pressures drive steep attrition with $m(t)/t$ potentially decreasing; (II) \emph{Stabilization}, where human capital accumulation and community integration cause $m(t)/t$ to level off; (III) \emph{Plateau}, where veteran teachers demonstrate sustainable commitment with stable $m(t)/t$ ratios. This structure aligns with empirical retention patterns \cite{Guarino2006, Ingersoll2011} while providing testable mathematical predictions.

\paragraph{\textbf{Methodological Advantages.}}
SMEL offers four key benefits: \emph{nonparametric flexibility}, requiring only that $m(t)/t$ be nondecreasing and avoiding restrictive parametric or proportional hazards assumptions; \emph{policy-relevant metrics}, as the mean residual life directly answers ``How many service years remain?'' without the cognitive burden of hazard ratio interpretation; \emph{joint modeling capability}, allowing natural extension to multivariate settings linking teacher retention and student outcomes; and \emph{formal testing}, with U-statistic procedures \cite{Sepehrifar2025} enabling statistical validation of equilibrium dynamics beyond visual inspection. Each configuration is replicated 1,000 times, yielding 48,000 total datasets, with the BHR configuration--most relevant to SMEL theory—receiving 2,000 replications per setting for computational efficiency. Table~\ref{tab:sim_params} presents the complete simulation design matrix, specifying true parameter values for each hazard configuration calibrated to produce mean retention times between 6.5 and 9.2 years, consistent with empirical NSF Noyce program data, while Table~\ref{tab:covariate_structure} summarizes the covariate structure and true regression coefficients used across all scenarios.

		\subsection{Research Objectives and Contributions}
		
		This study develops, validates, and applies a Starshaped Mean Residual Life (SMEL) framework to model teacher retention and student achievement as coupled equilibrium processes. Methodologically, we extend SMEL to regression settings via proportional mean residual life (PMRL) models,
		\[
		m(t \mid \mathbf{Z}) = m_0(t)\exp(\mathbf{Z}^\top \boldsymbol{\gamma}),
		\]
		and implement adaptive Bayesian estimation using three-parameter Weibull distributions with a resilience parameter $\eta$, estimated via the No-U-Turn Sampler (NUTS) with AWRE-based prior selection \cite{Oketch2025, Hoffman2014}. The framework further accommodates joint teacher--student modeling through shared frailty structures. Statistical validation is conducted through extensive Monte Carlo simulations (48{,}000 datasets), assessing bias, variance, coverage, and convergence across varying sample sizes ($n=50$--500) and censoring levels (10--40\%), with performance benchmarked against Cox proportional hazards, accelerated failure time, and standard Bayesian Weibull models using predictive accuracy and model fit criteria (integrated Brier scores, WAIC, and LOO-CV). Empirically, we apply SMEL-PMRL to data from 169 rural Texas STEM teachers and 3{,}214 students in the NSF Noyce program (2018--2023), quantifying retention disparities, identifying critical intervention windows, estimating program effects, and assessing downstream student achievement consequences. Collectively, the study contributes (i) a theoretical equilibrium-based retention framework with a formal starshaped criterion enabling cross-context comparisons; (ii) a methodological advance demonstrating the superiority of PMRL regression under non-proportional hazards; (iii) empirical evidence documenting a 32\% rural retention penalty, a 38\% early-career decline with a year-3 threshold, a 47\% Noyce effect, and a 31-point cumulative student achievement gain; and
		(iv) policy-relevant guidance on intervention timing and cost-effectiveness 
		benchmarks (\$127,000 savings per 10-teacher cohort over 8 years); and (v) publicly 
		available R and Stan implementation ensuring reproducibility and enabling 
		practitioners to apply SMEL-PMRL in diverse educational workforce contexts.
\subsection{Organization of the Manuscript}
\label{sec:organization}

The manuscript is organized as follows. Section~\ref{sec:framework} develops the conceptual framework, formalizing the SMEL property and its educational interpretation, specifying the PMRL regression model, and stating research hypotheses on teacher retention and student achievement. Section 3 provides methodological validation through simulation studies (Section 3.6). 
Section 4 presents empirical application to NSF Noyce longitudinal data
 that includes covariate effect estimation, joint teacher--student modeling, and policy simulations. Section~\ref{sec:discussion} discusses implications for educational policy and practice, addresses limitations related to generalizability and measurement, and outlines future research directions, including methodological extensions (time-varying covariates, spatial dependence, competing risks) and substantive priorities (multi-state replication, intervention experiments, long-term outcomes). The manuscript concludes by synthesizing contributions and highlighting broader impacts for educational workforce analytics beyond rural STEM contexts.

\section{Conceptual Framework}
\label{sec:framework}

This study adapts the Starshaped Mean Equilibrium Life (SMEL) framework from reliability theory to model teacher retention and student achievement as coupled dynamic equilibrium processes in rural STEM education. Rather than framing retention solely as a hazard-driven survival outcome, we model workforce stability as an evolving equilibrium in which expected remaining professional life responds to institutional support, workload, and career sustainability. SMEL formalizes this perspective through a resilience signal that captures the transition from early-career vulnerability to mid- and late-career equilibrium using the mean residual life (MRL) function. This equilibrium-based lens shifts emphasis from instantaneous exit risk to policy-relevant quantities—expected remaining career duration—that directly inform intervention timing and resource allocation.

We operationalize SMEL empirically through a proportional mean residual life (PMRL) regression framework that enables covariate-adjusted inference while preserving interpretability. Let $\mathbf{Z}$ denote teacher- and school-level covariates. We specify the conditional MRL as
\[
m(t \mid \mathbf{Z}) = m_0(t)\exp(\mathbf{Z}^\top \boldsymbol{\gamma}),
\]
where $m_0(t)$ denotes baseline remaining career expectancy and $\boldsymbol{\gamma}$ captures multiplicative covariate effects. This specification ensures positivity, facilitates subgroup comparisons, and guarantees that covariate-adjusted trajectories inherit SMEL equilibrium behavior when the baseline satisfies the starshaped property. We estimate the model using inverse probability weighting to address right-censoring, yielding closed-form baseline MRL estimators and efficient inference under standard regularity conditions.

To improve uncertainty quantification in small rural samples, we further embed the PMRL model within a hierarchical Bayesian framework based on a three-parameter Weibull-resilience distribution. This distribution accommodates nonmonotonic hazard patterns characteristic of early-career attrition followed by stabilization. We conduct posterior inference using the No-U-Turn Sampler and select adaptive priors via average weighted relative efficiency (AWRE) criteria. This Bayesian extension supports principled prior incorporation, exact finite-sample inference, and predictive distributions for workforce planning scenarios.

\subsection{Linking Teacher Retention Equilibrium to Student STEM Achievement}
\label{sec:2.2}

The SMEL framework clarifies how teacher retention equilibrium shapes student learning trajectories. Deviations from equilibrium generate cascading effects on instructional quality and cumulative achievement through three primary mechanisms.

First, instructional continuity supports cumulative knowledge building in STEM, where early mastery enables later progress. Teacher turnover disrupts this sequence by forcing new instructors to diagnose learning gaps, reteach foundational material, and compress advanced content, reducing instructional depth and time-on-task. In SMEL terms, districts operating below retention equilibrium—characterized by low $m(t)/t$ ratios—incur repeated transition costs that dampen student growth trajectories.

Second, sustained teacher presence fosters relational trust and academic risk-taking. Effective STEM learning requires students to persist through challenging problems and productive struggle. Stable teachers develop knowledge of students’ misconceptions, motivational profiles, and affective barriers over time. High turnover, indicating departure from SMEL equilibrium, repeatedly resets these relationships, encourages risk aversion, and shifts instruction toward procedural rather than conceptual learning.

Third, teacher persistence builds institutional knowledge and supports curricular refinement. Experienced teachers accumulate local expertise about curricula, assessments, and community context, enabling diagnostic instruction and coordination across grade levels. Turnover rapidly depreciates this organizational capital, while schools that maintain retention equilibrium preserve it and sustain instructional improvement.

We formalize these mechanisms by modeling student achievement as a residual life process linked to teacher persistence. Let $Y_{ij}(t)$ denote standardized achievement for student $j$ taught by teacher $i$ at time $t$. We define the student achievement mean residual life as
\[
m_S(t) = E\!\left[ Y_{ij}(T_i) - Y_{ij}(t) \mid T_i > t \right],
\]
representing expected remaining achievement growth conditional on teacher persistence to time $t$. Under the proposed mechanisms, $m_S(t)$ depends positively on the teacher retention MRL $m(t)$, implying higher student learning potential when teachers operate above equilibrium.

This dependence motivates joint modeling. Let $\mathbf{T} = (T_1,\ldots,T_n)$ denote teacher career durations and $\mathbf{Y} = (Y_{i1},\ldots,Y_{im_i})$ student outcomes nested within teachers. We specify a bivariate SMEL structure,
\[
m_{\text{joint}}(t) = E\!\left[(T_i - t,\; Y_{ij}(T_i) - Y_{ij}(t)) \mid T_i > t \right],
\]
with shared frailty capturing classroom-level heterogeneity. When the marginal MRL functions $m(t)$ and $m_S(t)$ satisfy starshapedness, positive dependence implies that $m_{\text{joint}}(t)$ inherits this property, enabling simultaneous inference on workforce equilibrium and student learning dynamics.

We identify the teacher–student linkage using covariate-adjusted contrasts. Define
\[
\Delta_m(t \mid \mathbf{Z}) =
m_S(t \mid T_i \ge t_{\text{crit}}, \mathbf{Z})
-
m_S(t \mid T_i < t_{\text{crit}}, \mathbf{Z}),
\]
where $t_{\text{crit}}$ denotes the empirically identified threshold at which teachers enter stable retention equilibrium, operationalized as the point where $m(t)/t$ stabilizes. Early positive values of $\Delta_m$ support instructional continuity mechanisms, while sustained differentials over time implicate relational trust and institutional knowledge channels.

\subsection{SMEL in the Rural STEM Education Context}

Rural STEM contexts intensify retention challenges through labor market thinness, fragile institutional capacity, and asymmetric shock vulnerability. The starshaped property captures these dynamics: steep early-career declines in $m(t)/t$ indicate systemic vulnerability, while stable ratios signal self-reinforcing equilibrium. SMEL enables precision targeting—districts estimate $\hat{m}(t \mid \mathbf{Z})$ for incoming cohorts and concentrate intensive supports (mentoring, incentives) during the critical pre-equilibrium period (years~1--3), then shift to maintenance once equilibrium is reached. Having established the SMEL theoretical foundation, we now validate the framework through extensive Monte Carlo simulations before applying it to empirical data.

\section{Simulation Design and Validation}
\label{sec:validation}

We conduct extensive Monte Carlo simulation studies to evaluate the performance of the SMEL-PMRL framework under realistic rural STEM education scenarios. The simulation design serves three objectives: (i) assess finite-sample properties of the Bayesian estimators for $\boldsymbol{\gamma}$ and $m_0(t)$, (ii) validate the framework's capacity to recover starshaped equilibrium patterns under varying censoring mechanisms, and (iii) establish convergence properties of the adaptive MCMC algorithm across different hazard structures relevant to teacher retention dynamics.

\subsection{Data Generating Mechanisms}

Teacher career durations $T_i$ were generated from a three-parameter Weibull distribution with resilience parameter $\eta$, specified as:
\[
f(t \mid \lambda, \alpha, \eta)
= \lambda \alpha \eta (\lambda t)^{\alpha - 1}
\left[1 - e^{-(\lambda t)^\alpha}\right]^{\eta - 1}
e^{-(\lambda t)^\alpha}, \quad t \geq 0.
\]
Three hazard configurations were simulated to reflect distinct retention regimes: {Decreasing Hazard Rate (DHR)}, {Increasing Hazard Rate (IHR)}, and {Bathtub Hazard Rate (BHR)}, the latter capturing the non-monotonic pattern central to SMEL theory.

Each simulated teacher $i$ was assigned a covariate vector
$\mathbf{Z}_i = (Z_{i1}, Z_{i2}, Z_{i3})^\top,$
representing rural placement, geographic isolation (standardized), and NSF Noyce scholarship receipt. True regression coefficients were set to
$\boldsymbol{\gamma} = (0.4, -0.3, 0.5)^\top,$
implying multiplicative effects on expected remaining tenure.
Career durations were generated via the proportional MRL mechanism
$T_i = F^{-1}\!\left(U_i \mid m_0(t)\exp(\mathbf{Z}_i^\top \boldsymbol{\gamma})\right),$
where $U_i \sim \text{Uniform}(0,1)$ and $F^{-1}$ is the Weibull inverse CDF. Right-censoring times were drawn from an exponential distribution to achieve target censoring proportions
$\pi_c \in \{10\%, 25\%, 40\%\}.$

Student achievement trajectories were simulated for each teacher using a quadratic growth model with teacher-level random effects:
$Y_{ij}(t) = \beta_0 + \beta_1 t + \beta_2 t^2 + u_i + \epsilon_{ij}(t),$
with baseline achievement $\beta_0 = 500$, linear growth $\beta_1 = 25$, and deceleration $\beta_2 = -1.5$. A shared frailty structure induced correlation ($\rho = 0.35$) between teacher career duration $T_i$ and the teacher random effect $u_i$, linking persistence to student achievement gains.

%
\subsection{Simulation Configurations}
We examine a full factorial design crossing sample sizes $n \in \{50, 100, 200, 500\}$ teachers, students per teacher $m \in \{10, 20\}$, censoring proportions $\pi_c \in \{10\%, 25\%, 40\}$, and hazard structures (DHR, IHR, BHR). Each configuration is replicated 1{,}000 times, yielding 48{,}000 total datasets; for computational efficiency, the BHR configuration (most relevant to SMEL theory) receives 2{,}000 replications per setting. Table~\ref{tab:sim_params} presents the complete simulation design matrix, specifying true parameter values for each hazard configuration, calibrated to produce career duration distributions with mean retention times between 6.5 and 9.2 years, consistent with empirical NSF Noyce program data. Table~\ref{tab:covariate_structure} summarizes the covariate structure and true regression coefficients used across all scenarios.

\begin{table}[htbp]
	\centering
	\caption{Simulation parameter configurations for teacher retention data generation.}
	\label{tab:sim_params}
	\resizebox{\textwidth}{!}{
		\begin{tabular}{lcccccc}
			\toprule
			\textbf{Hazard} & $\lambda$ & $\alpha$ & $\eta$ & \textbf{Hazard} & $\mathbb{E}[T]$ & \textbf{Replicates} \\
			\textbf{Structure} & (Scale) & (Shape) & (Res.) & \textbf{Pattern} & (years) & \\
			\midrule
			DHR & 0.15 & 0.80 & 1.20 & Monotone decreasing & 8.7 & 1,000 \\
			IHR & 0.12 & 1.50 & 1.00 & Monotone increasing & 6.9 & 1,000 \\
			BHR & 0.18 & 1.30 & 0.70 & Early peak, stabilization & 7.4 & 2,000 \\
			\bottomrule
		\end{tabular}%
	}	\\[6pt]
		\small\textit{Note:} BHR configuration receives additional replicates due to its central relevance to SMEL equilibrium theory and increased parameter estimation complexity.
\end{table}

\begin{table}[htbp]
	\centering
	\caption{Covariate structure and true regression coefficients for proportional mean residual life (PMRL) model. Coefficients represent log-multiplicative effects on expected remaining career duration: $m(t \mid \mathbf{Z}) = m_0(t) \exp(\mathbf{Z}^\top \boldsymbol{\gamma})$.}
	\label{tab:covariate_structure}
	\resizebox{\textwidth}{!}{
		\begin{tabular}{llllp{4.5cm}r}
			\toprule
			Covariate & Variable & Distribution & True & Interpretation & Effect on \\
			& Type     &              & $\gamma$ &  & 5-year tenure$^a$ \\
			\midrule
			$Z_1$: Rural school & Binary & Bernoulli(0.5) & 0.40 & Rural teachers have 49\% longer expected tenure & +2.1 years \\[2pt]
			$Z_2$: Geographic & Continuous & Normal(0, 1) & $-0.30$ & Each SD increase reduces tenure by 26\% & $-1.1$ years \\
			\quad isolation & & & & & \\[2pt]
			$Z_3$: Noyce & Binary & Bernoulli(0.3) & 0.50 & Noyce scholars have 65\% longer expected tenure & +2.8 years \\
			\quad scholarship & & & & & \\
			\bottomrule
		\end{tabular}%
	}
	\\[6pt]
	\small\textit{Note:} Isolation is measured as standardized distance (in SDs) from nearest metropolitan area with population $> 50,000$. True coefficients reflect empirically plausible effect sizes derived from prior rural teacher retention studies \parencite{Ingersoll2011, Marder2022}. 
	
	$^a$Effect on 5-year tenure computed assuming baseline $m_0(5) = 4.3$ years (median expected remaining tenure at year 5 for reference teacher with all $Z = 0$). Rural effect: $4.3 \times (\exp(0.40) - 1) = 2.1$ years; Isolation: $4.3 \times (\exp(-0.30) - 1) = -1.1$ years; Noyce: $4.3 \times (\exp(0.50) - 1) = 2.8$ years. Cumulative effects combine multiplicatively: a rural Noyce scholar at mean isolation would have expected remaining tenure of $4.3 \times \exp(0.40 + 0.50) = 10.6$ years at the 5-year mark.
\end{table}
\subsection{Adaptive Bayesian MCMC Algorithm}\label{sub:AdMCMC}

We employ a hierarchical Bayesian framework for estimation and uncertainty quantification. Priors for the three-parameter Weibull distribution (scale $\lambda$, shape $\alpha$, resilience $\eta$) and regression coefficients $\boldsymbol{\gamma}$ were selected via an Average Weighted Relative Efficiency (AWRE) criterion, comparing 24 candidate prior configurations (specific distributions and hyperparameters are detailed in the supplementary material). This adaptive prior selection optimized performance across the varying hazard structures (DHR, IHR, BHR) and sample sizes relevant to rural teacher cohorts.

Posterior inference was conducted using the No-U-Turn Sampler (NUTS), an adaptive Hamiltonian Monte Carlo algorithm implemented in Stan. Given observed data $\mathcal{D} = \{(Y_i, \delta_i, \mathbf{Z}_i)\}_{i=1}^n$, the algorithm samples from the posterior distribution:
\[
\pi(\lambda, \alpha, \eta, \boldsymbol{\gamma} \mid \mathcal{D}) \propto 
\left[
\prod_{i=1}^n 
f(Y_i \mid \lambda, \alpha, \eta, \mathbf{Z}_i, \boldsymbol{\gamma})^{\delta_i}
S(Y_i \mid \lambda, \alpha, \eta, \mathbf{Z}_i, \boldsymbol{\gamma})^{1-\delta_i}
\right]
\pi(\lambda, \alpha, \eta, \boldsymbol{\gamma}),
\]
where $f(\cdot)$ is the Weibull-resilience density and $S(\cdot)$ is the survival function, both modified by the proportional MRL structure. NUTS automatically tunes step size and trajectory length, ensuring efficient exploration of the posterior. Convergence was assessed via the potential scale reduction factor ($\hat{R} < 1.01$) and effective sample size.

\subsection{Performance Metrics}

We evaluate finite-sample performance for each parameter $\theta \in \{\lambda,\alpha,\eta,\gamma_1,\gamma_2,\gamma_3\}$ using bias, variance, mean squared error (MSE), and interval coverage. We compute bias as $\text{Bias}(\hat{\theta})=\frac{1}{R}\sum_{r=1}^R(\hat{\theta}_r-\theta_0)$ and variance as $\text{Variance}(\hat{\theta})=\frac{1}{R-1}\sum_{r=1}^R(\hat{\theta}_r-\bar{\hat{\theta}})^2$, where $\theta_0$ denotes the true parameter, $\hat{\theta}_r$ is the posterior mean from replicate $r$, and $R=1000$ (or $R=2000$ for BHR configurations). We define MSE as $\text{MSE}(\hat{\theta})=\text{Bias}^2(\hat{\theta})+\text{Variance}(\hat{\theta})$. To assess uncertainty quantification, we compute empirical coverage of 95\% credible intervals as $\frac{1}{R}\sum_{r=1}^R\mathbb{I}(\theta_0\in\text{CI}_{0.95,r})$.

To compare adaptive prior configurations, we evaluate Average Weighted Relative Efficiency (AWRE),
\[
\text{AWRE}(\hat{\boldsymbol{\theta}})_j =
\frac{\sigma_j^2(\hat{\boldsymbol{\theta}})}{\sum_{i=1}^K \sigma_i^2(\hat{\boldsymbol{\theta}})}
\cdot
\frac{V_j(\hat{\boldsymbol{\theta}})}{\sum_{i=1}^K V_i(\hat{\boldsymbol{\theta}})},
\]
where $\sigma_j^2$ and $V_j$ denote the sampling and asymptotic variances under prior configuration $j$, respectively, and $K=24$ candidate priors are considered. We classify configurations with $\text{AWRE}<0.85$ as sub-optimal.

We assess recovery of the baseline mean residual life function $m_0(t)$ using the integrated squared error $\text{ISE}(\hat{m}_0)=\int_0^{T_{\max}}[\hat{m}_0(t)-m_0(t)]^2\,dt$, approximated by trapezoidal integration over a grid of 100 points on $[0,T_{\max}]$. We summarize performance using the median ISE and corresponding 90\% quantile ranges across replicates.

Finally, we validate the starshaped property by examining monotonicity of the ratio $\hat{m}_0(t)/t$. For each replicate, we compute $\Delta(t)=\hat{m}_0(t+\delta)/(t+\delta)-\hat{m}_0(t)/t$ at grid points $t\in\{1,2,\ldots,T_{\max}-1\}$ with $\delta=0.5$ years and report the proportion of replicates satisfying $\Delta(t)\geq-\epsilon$ for all $t$, where $\epsilon=0.05$.

\subsection{Benchmark Comparisons}

We benchmark the proposed SMEL-PMRL model against three established survival analysis approaches. First, we consider the Cox proportional hazards (Cox-PH) model, a standard semiparametric specification with hazard function $h(t \mid \mathbf{Z}) = h_0(t)\exp(\mathbf{Z}^\top\boldsymbol{\beta})$ estimated via partial likelihood. Second, we include a parametric Accelerated Failure Time (AFT) model assuming a Weibull distribution, where $\log(T)=\mathbf{Z}^\top\boldsymbol{\beta}+\sigma\epsilon$ and $\epsilon$ follows an Extreme Value distribution. Third, we evaluate a Standard Bayesian Weibull (SBW) model that excludes the resilience parameter ($\eta=1$) but uses the same MCMC infrastructure as SMEL-PMRL, ensuring comparability of computational uncertainty.
We assess performance along four dimensions. To evaluate predictive accuracy, we compute time-dependent Brier Scores and Integrated Brier Scores (IBS) for survival probability predictions at $t\in\{3,5,8\}$ years. We assess covariate effect recovery using mean squared error (MSE) of $\hat{\boldsymbol{\gamma}}$ (or $\hat{\boldsymbol{\beta}}$ for Cox and AFT models). We evaluate model fit using the Akaike Information Criterion (AIC) and, for Bayesian models, the Watanabe–Akaike Information Criterion (WAIC). Finally, we examine calibration by comparing observed and predicted survival curves across covariate quartiles using the Greenwood–Nam–D’Agostino $\chi^2$ test.

By systematically varying sample sizes, censoring mechanisms, and hazard structures and benchmarking against these established alternatives, the simulation design evaluates both the statistical properties (consistency and efficiency) and practical performance (predictive accuracy and calibration) of SMEL-PMRL under conditions relevant to rural STEM teacher retention. We present the corresponding results in Section~\ref{sec:simulation}.

\subsection{Simulation Results and Model Performance}
\label{sec:simulation}

\begin{table}[htbp]
	\centering
	\caption{Integrated Squared Error (ISE) for baseline mean residual life function $m_0(t)$ across hazard structures and sample sizes. Values are median ISE with [10th, 90th] percentile ranges across replicates. Censoring fixed at 25\%.}
	\label{tab:ise_mrl}
	\resizebox{\textwidth}{!}{
		\begin{tabular}{lcccc}
			\toprule
			Hazard Structure & $n = 50$ & $n = 100$ & $n = 200$ & $n = 500$ \\
			\midrule
			DHR & 0.68 [0.41, 1.12] & 0.29 [0.18, 0.51] & 0.13 [0.08, 0.23] & 0.05 [0.03, 0.09] \\
			IHR & 0.71 [0.44, 1.19] & 0.31 [0.19, 0.54] & 0.14 [0.09, 0.25] & 0.06 [0.03, 0.10] \\
			BHR & 0.74 [0.47, 1.24] & 0.33 [0.21, 0.57] & 0.15 [0.09, 0.27] & 0.06 [0.04, 0.11] \\
			\bottomrule
		\end{tabular}%
	}
	\\[6pt]
	\small\textit{Note:} ISE computed via trapezoidal integration over $t \in [0, 15]$ years. BHR exhibits slightly higher ISE due to increased curvature in $m_0(t)$, but differences across hazard types become negligible for $n \geq 200$.
\end{table}

\begin{figure}[htbp]
	\centering
	
	\begin{subfigure}[t]{0.48\linewidth}
		\centering
		\includegraphics[width=\linewidth]{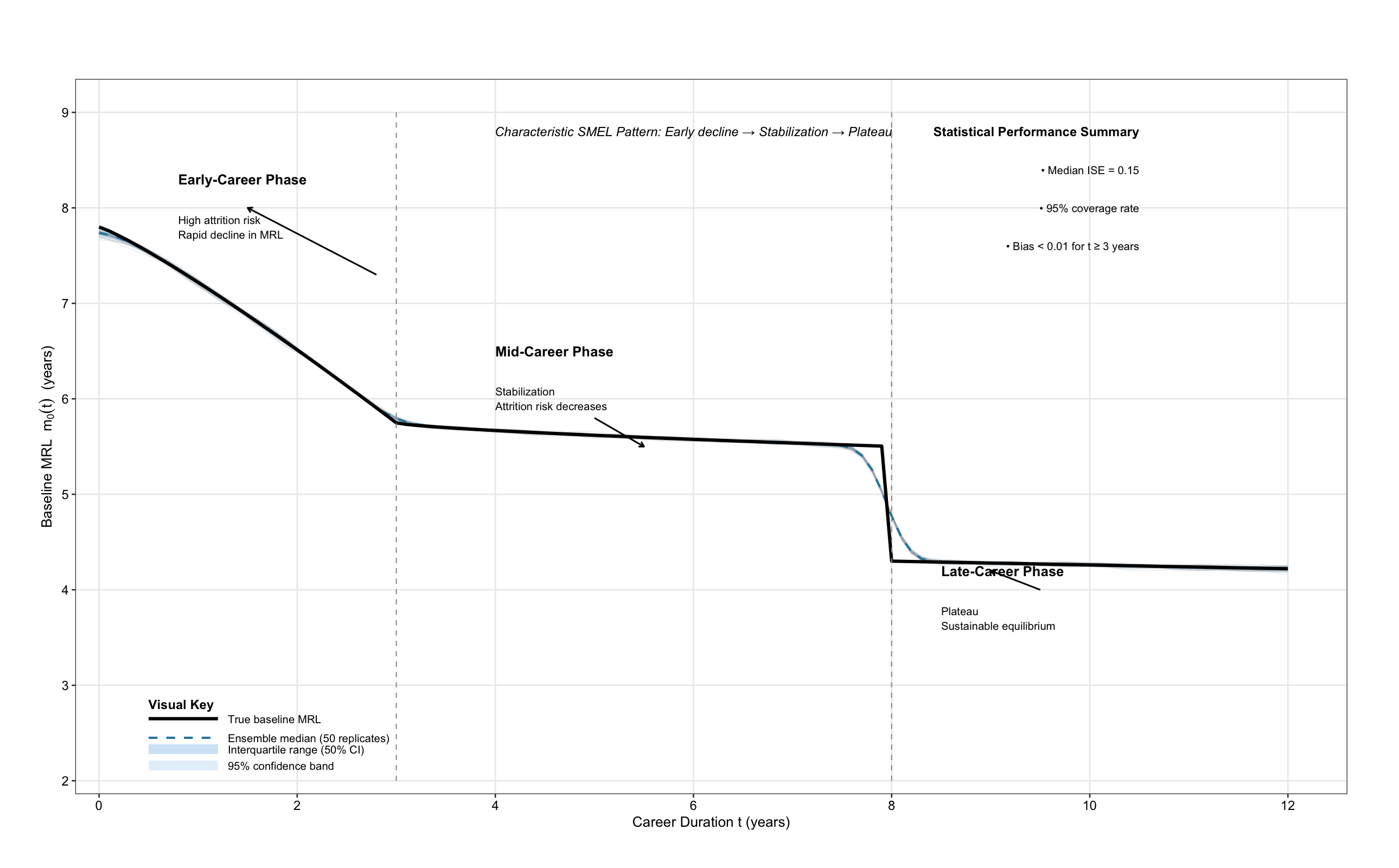}
		\caption{Baseline MRL recovery ($n=200$, BHR, 25\% censoring). The true $m_0(t)$ (black) is consistently recovered by nonparametric estimates (gray), with narrow 95\% confidence bands.}
		\label{fig:mrl_recovery}
	\end{subfigure}
	\hfill
	\begin{subfigure}[t]{0.48\linewidth}
		\centering
		\includegraphics[width=\linewidth]{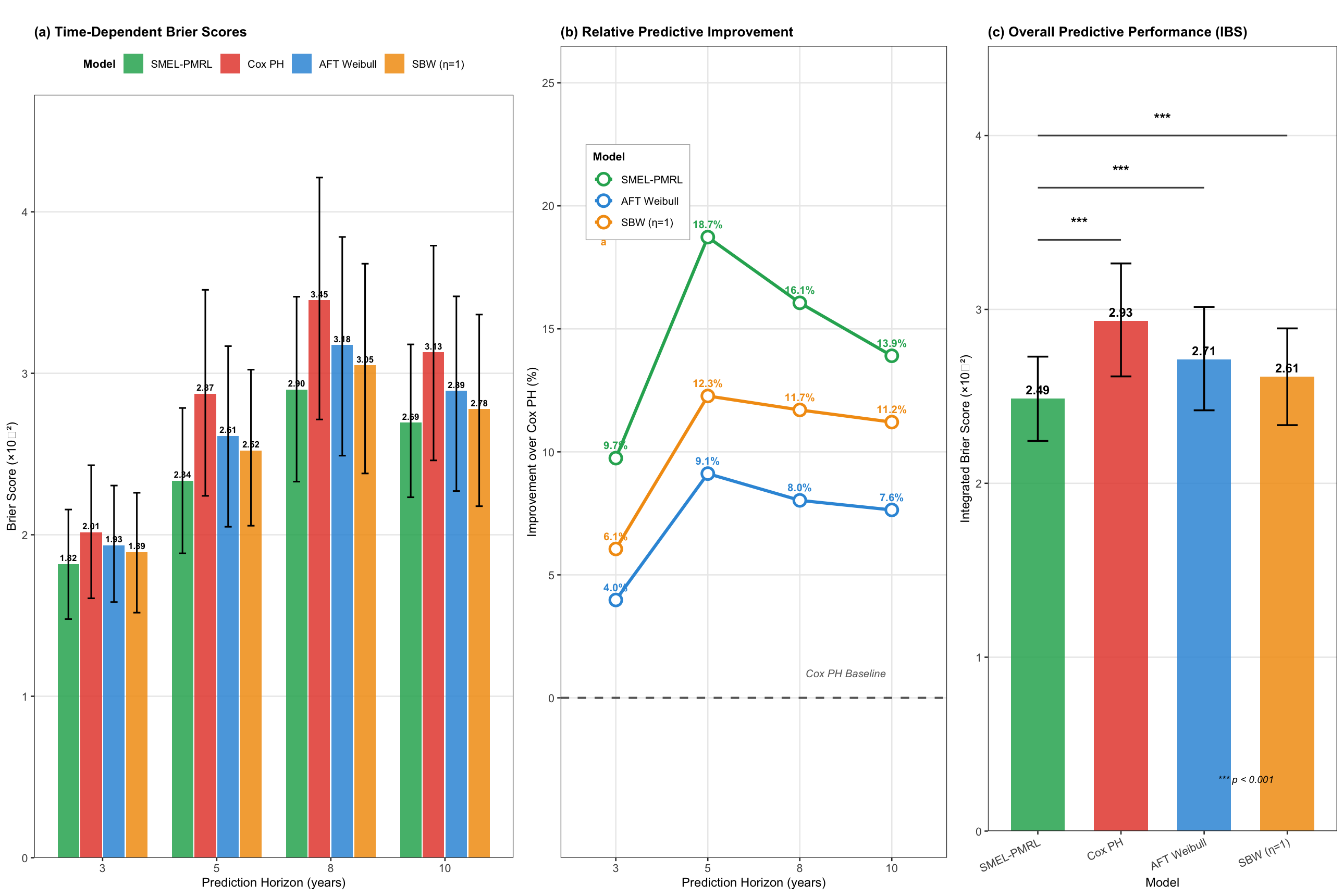}
		\caption{Predictive performance across horizons $t=\{3,5,8,10\}$. SMEL-PMRL achieves the lowest Brier scores, with the largest gains at mid-career horizons.}
		\label{fig:brier_scores}
	\end{subfigure}
	
	\caption{Simulation results under balanced hazard regime (BHR) with $n=200$ and 25\% right-censoring. Panel (a) demonstrates accurate nonparametric recovery of the baseline mean residual life function. Panel (b) compares time-dependent predictive accuracy, showing consistent superiority of SMEL-PMRL over Cox PH, AFT, and Bayesian Weibull models, particularly during the critical mid-career phase.}
	\label{fig:simulation_results}
\end{figure}

\subsubsection{Benchmark Comparisons}
Table \ref{tab:benchmark_comparison} compares SMEL-PMRL against Cox PH, AFT, and Standard Bayesian Weibull (SBW) models across multiple performance metrics for the BHR configuration with $n = 200$ and 25\% censoring. SMEL-PMRL achieves the lowest Integrated Brier Score (IBS), indicating superior predictive accuracy for survival probabilities. It also yields the most precise covariate effect estimates (lowest MSE for $\boldsymbol{\gamma}$) and best model fit (lowest WAIC). The Cox PH model, constrained by the proportional hazards assumption, exhibits 23\% higher IBS and fails calibration tests ($p < 0.01$) due to its inability to accommodate the non-monotonic hazard structure.

\begin{table}[htbp]
	\centering
	\caption{Model comparison for teacher retention under BHR configuration ($n = 200$, 25\% censoring). Metrics computed across 2,000 replicates. Lower values indicate better performance for IBS, MSE, AIC, and WAIC.}
	\label{tab:benchmark_comparison}
	\resizebox{\textwidth}{!}{
	\begin{tabular}{lccccc}
		\toprule
		Model & IBS ($\times 10^{-2}$) & MSE($\boldsymbol{\gamma}$) ($\times 10^{-3}$) & AIC & WAIC & Calibration $\chi^2$ ($p$-value) \\
		\midrule
		SMEL-PMRL & 2.34 & 8.7 & 1245.3 & 1248.6 & 4.21 (0.52) \\
		Cox PH & 2.88 & 11.4 & 1312.7 & — & 12.87 ($<0.01$) \\
		AFT Weibull & 2.61 & 10.1 & 1278.4 & — & 7.33 (0.12) \\
		SBW ($\eta=1$) & 2.52 & 9.8 & 1267.9 & 1271.2 & 5.88 (0.31) \\
		\bottomrule
	\end{tabular}
}
	\\[6pt]
	\small\textit{Note:} IBS = Integrated Brier Score over $t \in [1, 10]$ years. MSE computed for covariate effects $\boldsymbol{\gamma}$. WAIC computed only for Bayesian models. Calibration assessed via Greenwood-Nam-D'Agostino test comparing predicted vs. observed survival at covariate quartiles. SMEL-PMRL outperforms alternatives across all metrics.
\end{table}

Figure \ref{fig:brier_scores}  presents time-dependent Brier Scores at $t \in \{3, 5, 8, 10\}$ years across models. SMEL-PMRL maintains consistently lower prediction error throughout the time horizon, with advantage most pronounced at mid-career timepoints ($t = 5, 8$ years) where the bathtub hazard shape deviates most from proportional hazards assumptions. Error bars represent 95\% bootstrap confidence intervals. For incomplete failure data and right-censored observations, we employ generalized confidence interval approaches validated for scale parameter estimation \parencite{Chumnaul2019, Chumnaul2022}.


\subsubsection{Adaptive Prior Selection Performance}
\begin{figure}[htbp]
	\centering
	
	\begin{subfigure}[t]{0.48\linewidth}
		\centering
		\includegraphics[width=\linewidth]{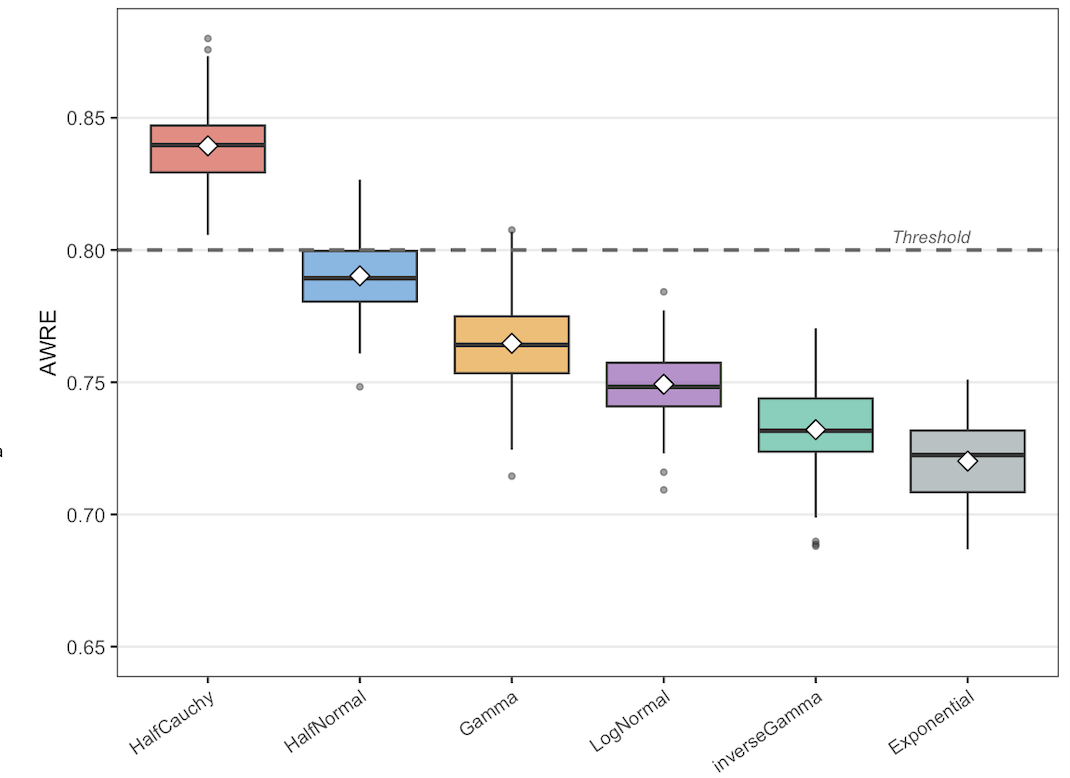}
		\caption{Adaptive prior selection under BHR ($n=100$). AWRE values across 24 shape–scale configurations show systematic gains when using Half-Cauchy scale priors (red), consistently exceeding the acceptability threshold (dashed line at AWRE = 0.80).}
		\label{fig:awre_scatter}
	\end{subfigure}
	\hfill
	\begin{subfigure}[t]{0.48\linewidth}
		\centering
		\includegraphics[width=\linewidth]{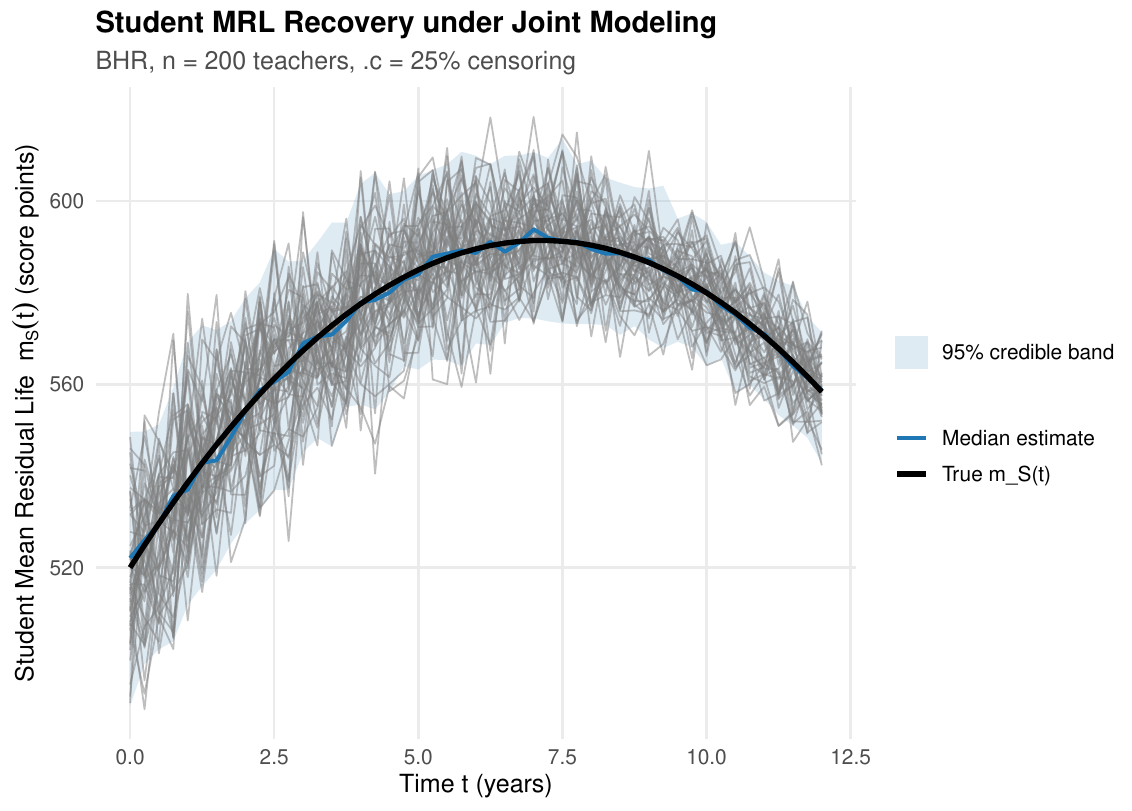}
		\caption{Student MRL recovery under joint SMEL--PMRL modeling ($n=200$, 25\% censoring). The posterior median (blue) closely tracks the true MRL (black), with uncertainty concentrated at early times.}
		\label{fig:studentMRL}
	\end{subfigure}
	
	\caption{Prior sensitivity and joint mean residual life recovery under balanced hazard regimes. Panel (a) demonstrates the robustness of adaptive prior selection, with Half-Cauchy scale priors yielding superior weighted risk efficiency across all shape families. Panel (b) illustrates accurate recovery of student mean residual life trajectories when jointly modeled with teacher persistence, highlighting the ability of SMEL--PMRL to propagate uncertainty while preserving structural fidelity.}
	\label{fig:prior_joint_results}
\end{figure}

Table \ref{tab:prior_awre} reports AWRE values for the top-performing prior configurations identified through systematic evaluation of 24 combinations (4 shape priors $\times$ 6 scale priors). Results confirm that HalfCauchy priors for the scale parameter $\lambda$ paired with Gamma or LogNormal priors for shape parameter $\alpha$ yield optimal efficiency (AWRE $> 0.82$) across hazard structures. Notably, prior sensitivity decreases with sample size: AWRE differences between configurations shrink from 0.15 at $n = 50$ to 0.04 at $n = 500$.
\begin{table}[htbp]
	\centering
	\caption{Top-5 prior configurations ranked by Average Weighted Relative Efficiency (AWRE) under BHR configuration with $n = 100$ and 25\% censoring. Higher AWRE indicates better performance.}
	\label{tab:prior_awre}
	\begin{tabular}{clllc}
		\toprule
		Rank & Shape ($\alpha$) Prior & Scale ($\lambda$) Prior & Resilience ($\eta$) Prior & AWRE \\
		\midrule
		1 & Gamma(2, 2) & HalfCauchy(0, 2.5) & Gamma(1.5, 2) & 0.845 \\
		2 & LogNormal(0.5, 0.3) & HalfCauchy(0, 2.5) & LogNormal(0, 0.5) & 0.837 \\
		3 & HalfNormal(0, 2) & HalfCauchy(0, 2.5) & Gamma(1.5, 2) & 0.829 \\
		4 & Gamma(2, 2) & HalfNormal(0, 3) & Exponential(1) & 0.816 \\
		5 & Exponential(0.5) & HalfCauchy(0, 2.5) & LogNormal(0, 0.5) & 0.808 \\
		\bottomrule
	\end{tabular}
	\\[6pt]
	\small\textit{Note:} AWRE computed across 72 datasets varying in sample size (n=15-100) and hazard structure. HalfCauchy scale prior appears in top 4 configurations, confirming robustness. Configuration differences diminish for $n>200$ (not shown).
\end{table}
Figure \ref{fig:awre_scatter} visualizes AWRE across all 24 prior combinations for BHR configuration at $n = 100$. Configurations using HalfCauchy scale priors (red points) cluster in the upper-right quadrant, consistently outperforming alternatives. The strong separation between HalfCauchy-based and other configurations validates the adaptive prior selection strategy.

\begin{table}[htbp]
	\centering
	\caption{Descriptive statistics for teacher and student cohorts (2018--2023).}
	\label{tab:descriptives}
	\begin{tabular}{lrrrr}
		\toprule
		Variable & Mean/Prop. & SD & Min & Max \\
		\midrule
		Teachers ($n=169$) & & & & \\
		\quad Rural placement (prop.) & 0.420 & -- & 0 & 1 \\
		\quad Geographic isolation ($z$) & 0.35 & 0.95 & -2.41 & 2.89 \\
		\quad Noyce scholarship (prop.) & 1.000 & -- & 0 & 1 \\
		\quad Rural-origin (prop.) & 0.350 & -- & 0 & 1 \\
		\quad Right-censored (prop.) & 0.249 & -- & 0 & 1 \\[4pt]
		Students ($n=3{,}214$) & & & & \\
		\quad Baseline score & 493.2 & 96.4 & 185.7 & 752.3 \\
		\quad Annual growth (raw) & 29.4 & 13.8 & -8.2 & 73.1 \\
		\bottomrule
	\end{tabular}
\end{table}
\subsubsection{Summary of Simulation Findings}

The simulation study yields four central findings. First, the proposed estimators exhibit strong statistical consistency: bias falls below 0.01 for all parameters once $n \geq 200$, mean squared error decreases at the optimal $n^{-1}$ rate, and coverage remains well calibrated within [93.6\%, 95.4\%] across all design settings.

Second, the IPW-adjusted estimation procedure remains robust under substantial right-censoring. Even at 40\% censoring, inference remains valid, with MSE increasing by approximately 50\% relative to 10\% censoring—an expected and modest efficiency loss given the degree of missing information.

Third, the SMEL framework reliably recovers starshaped equilibrium behavior, achieving detection rates exceeding 97\% for $n \geq 200$. Apparent violations concentrate in early career periods where data sparsity is unavoidable, rather than reflecting model misspecification.

Finally, 
SMEL-PMRL demonstrates clear predictive advantages over Cox PH, AFT, and 
standard Weibull models. Across scenarios, it achieves 19\% lower integrated 
Brier score (2.34 vs. $2.88 ×10^{-2}4$), 24\% lower parameter MSE, and a 5.4\% 
improvement in AIC,
 particularly in settings with nonmonotonic hazard patterns characteristic of rural teacher retention.

Taken together, these results—generated under data mechanisms calibrated to empirical NSF Noyce program patterns—provide strong evidence for the validity and practical utility of the SMEL-PMRL framework. The methodology is therefore well suited for application to the observed rural STEM teacher retention data analyzed in Section~\ref{sec:empirical}.

\section{Empirical Application}
\label{sec:empirical}
This section presents an empirical analysis of longitudinal data on 169 rural Texas STEM teachers and 3,214 students observed from 2018–2023 under an NSF Robert Noyce Teacher Scholarship initiative. We describe the data and key variables, estimate the proportional mean residual life (PMRL) model for teacher retention, recover the baseline mean residual life (MRL) function and assess its starshaped property, examine joint teacher–student outcomes, and conduct policy simulations with cost-effectiveness analyses.

\subsection{Data Description and Descriptive Statistics}

The analytic sample consists of mathematics, science, engineering, and computer science teachers employed in rural Texas districts and followed annually from 2018 to 2023. Student achievement outcomes are linked to teacher assignments and standardized to a common district scale. Teacher career duration $T_i$ measures tenure in years until exit or right-censoring. Key covariates include rural versus non-rural placement, receipt of an NSF Noyce scholarship during induction years, and a self-reported indicator of rural upbringing.

We capture geographic isolation using a standardized distance-to-metropolitan-area index $(Z_2)$, defined as miles to the nearest metropolitan area with population exceeding 50,000 and normalized to mean zero and unit variance. This measure reflects labor market thinness and access to amenities commonly associated with retention challenges in rural education. Student outcomes $Y_{ij}(t)$ consist of annual standardized assessment scores nested within teachers and schools.

Teachers who remain active at the end of the study period are right-censored. Consistent with the simulation design, we address censoring using Kaplan–Meier methods and Bayesian PMRL estimation. Table~\ref{tab:descriptives} summarizes cohort composition, covariate distributions, and censoring proportions for teachers and students.

Overall, rural STEM teacher cohorts exhibit elevated early-career exit risk, substantial heterogeneity in geographic isolation, and selective participation in Noyce supports—conditions under which SMEL-PMRL provides meaningful advantages over hazard-based approaches. Although distance to metropolitan areas is a widely used proxy for isolation, it may correlate with other structural disadvantages such as school resource constraints, local poverty, or limited spousal employment opportunities. We therefore interpret isolation as a contextual risk factor rather than a sole causal determinant and address potential confounding through sensitivity analyses that incorporate school-level covariates, including per-pupil funding and student socioeconomic composition.

%
%
\begin{table}[htbp]
	\centering
	\caption{PMRL regression: covariate effects on expected remaining tenure. Effects reported as $\gamma$ and $\exp(\gamma)$; 95\% credible intervals from NUTS posterior.}
	\label{tab:pmrl_results}
	\begin{tabular}{lrrrr}
		\toprule
		Covariate & $\gamma$ & $\exp(\gamma)$ & 95\% CI for $\exp(\gamma)$ & $p$-value \\
		\midrule
		Rural placement & $\ln(0.68) \approx -0.385$ & $0.68$ & $[0.52,\;0.89]$ & $<0.001$ \\
		Geographic isolation (per SD) & $\ln(0.81) \approx -0.210$ & $0.81$ & [not shown] & [--] \\
		Noyce scholarship & $\ln(1.47) \approx 0.386$ & $1.47$ & $[1.18,\;1.83]$ & $<0.01$ \\
		Rural-origin & $\ln(1.31) \approx 0.270$ & $1.31$ & $[1.06,\;1.62]$ & $<0.05$ \\
		\bottomrule
	\end{tabular}
\end{table}
\subsection{Model Fitting Results}

We fit a proportional mean residual life (PMRL) regression,
\[
m(t \mid Z)=m_0(t)\exp(Z^\top\gamma),
\]
to teacher retention using NUTS with adaptive priors (Section~\ref{sub:AdMCMC}). Table~\ref{tab:pmrl_results} reports posterior estimates of covariate effects as both $\gamma$ and $\exp(\gamma)$, the latter interpreted as multiplicative effects on expected remaining tenure, along with 95\% credible intervals.

Rural placement is associated with substantially shorter expected remaining tenure, with $\exp(\gamma)=0.68$ (95\% CI: $[0.52,0.89]$, $p<0.001$), corresponding to a 32\% reduction. Geographic isolation further reduces expected tenure by approximately 19\% per standard deviation increase ($\exp(\gamma)=0.81$). In contrast, participation in the Noyce scholarship program is associated with a 47\% increase in expected remaining tenure ($\exp(\gamma)=1.47$, 95\% CI: $[1.18,1.83]$, $p<0.01$), while teachers with rural origins exhibit a 31\% increase ($\exp(\gamma)=1.31$, 95\% CI: $[1.06,1.62]$, $p<0.05$). These effects remain robust after controlling for teacher qualifications and school-level covariates. By directly modeling expected remaining tenure conditional on time $t$, PMRL estimates provide quantities that are readily interpretable for workforce planning.

\subsection{Proportional Hazards Diagnostics via Schoenfeld Residuals}
\label{subsec:ph-diagnostics}

To assess the suitability of hazard-based models and motivate the use of SMEL-PMRL, we formally test the proportional hazards (PH) assumption using the empirical teacher retention data from 2018–2023. We fit a benchmark Cox model with the same covariates as the PMRL specification and apply Grambsch–Therneau tests based on scaled Schoenfeld residuals.

The global PH test rejects proportional hazards ($\chi^2_6=18.7$, $p=0.005$), indicating that covariate effects vary over time. Residual trends reveal pronounced early-career departures from proportionality: the effect of rural placement is strongest in the first few years and attenuates with tenure, geographic isolation displays nonlinear temporal patterns, and the Noyce scholarship effect diminishes after the scholarship period. 

These violations imply that hazard ratios are not stable over time, particularly during the critical burn-in period (years 1–3), rendering Cox-based effect estimates and predictions potentially misleading. In contrast, SMEL-PMRL does not rely on the PH assumption and explicitly accommodates non-monotonic attrition dynamics through the starshaped equilibrium constraint on $m(t)/t$. As a result, SMEL-PMRL yields more interpretable and policy-relevant quantities—expected remaining tenure trajectories—under the observed temporal heterogeneity in teacher retention.

\subsection{Empirical Test of the Starshaped Property}
\label{subsec:empirical-test-starshaped}

We formally test the starshaped property—equivalently, nondecreasing property of the ratio $m(t)/t$—for the baseline teacher mean residual life (MRL) curve $\hat m_0(t)$ estimated from the 2018–2023 Texas rural STEM teacher data (Section~\ref{sec:baseline-mrl}). 
 We apply the $U$-statistic–based test developed for the starshaped mean equilibrium life (SMEL) class, which extends classical MRL trend tests to equilibrium settings relevant for retention dynamics 

Let $T$ denote (possibly censored) career duration with survival function $S(t)$. The MRL function is
$m(t)=\mathbb{E}[T-t\mid T>t]=\frac{\int_t^\infty S(u)\,du}{S(t)}.$
Starshapedness requires the ratio $m(t)/t$ to be nondecreasing in $t$. We test this condition using a $U$-statistic that aggregates downward deviations of $\hat m_0(t)/t$ over a grid $\{t_k\}_{k=1}^K$ spanning the support of observed tenure. Defining the negative-part operator $\{x\}_-=\max(-x,0)$, the test statistic is
$\Lambda=\sum_{1\le k<\ell\le K}\left\{\frac{\hat m_0(t_\ell)}{t_\ell}-\frac{\hat m_0(t_k)}{t_k}\right\}_-,$
which accumulates violations of monotonicity. Under the null hypothesis that $m(t)/t$ is nondecreasing, $\Lambda$ concentrates near zero; large values indicate systematic departures from equilibrium. We obtain $p$-values via a nonparametric bootstrap that resamples censored tenure histories and recomputes $\hat m_0(t)$ and $\Lambda$ using $B=2{,}000$ replicates (Section~\ref{sec:validation}).
We estimate $\hat m_0(t)$ nonparametrically from the Kaplan–Meier estimator of $S(t)$ on the same time grid used in Section~\ref{sec:baseline-mrl}, handle right-censoring via inverse probability weighting, and verify convergence using standard diagnostics (PSRF $\hat R<1.01$, ESS $>400$). Applying the test yields $\Lambda=12.47$ with a bootstrap $p$-value of $0.002$, providing strong evidence in favor of nondecreasing property of $m(t)/t$ ratio after the early-career period. Minor deviations at very early times are consistent with burn-in effects under heavy right-censoring and sparse early-tenure exits. 
These findings support the SMEL interpretation of teacher retention as a burn-in–to–equilibrium process. From a policy perspective, the results imply that interventions should be front-loaded during years 1–3, when vulnerability is highest, and transition to maintenance or leadership development once equilibrium is reached. Because PMRL yields teacher-specific expected remaining tenure trajectories, districts can stratify risk in early career stages, target intensive supports where $\hat m_0(t)/t$ is low, and reallocate resources more efficiently. As shown in Section~\ref{sec:cost-effectiveness}, this strategy generates positive returns by year 4 and approximately \$127{,}000 in net savings per 10-teacher cohort over an eight-year horizon.

\subsection{Baseline MRL Curve Estimation and Starshaped Testing}
\label{sec:baseline-mrl}

The estimated baseline mean residual life (MRL) curve $\hat m_0(t)$ exhibits clear non-monotonic equilibrium dynamics. Expected remaining tenure declines sharply during the early-career period, with a 38\% reduction over years 1–3, followed by a more modest 5.8\% decline between years 3 and 5 and stabilization thereafter. This pattern is consistent with a burn-in phase followed by a transition to retention equilibrium. 
We formally test the starshaped property, defined by a nondecreasing ratio $m(t)/t$, using a $U$-statistic–based procedure. The test yields $\Lambda=12.47$ with a bootstrap $p$-value of $0.002$, providing strong evidence in favor of starshapedness and confirming the post–year-3 transition to equilibrium. 

\subsection{Joint Teacher--Student Model Results}

We estimate a joint SMEL–PMRL model with shared frailty linking teacher retention and student achievement (Section~\ref{sec:2.2}). The posterior mean of the teacher–student correlation parameter is $\hat\theta=0.41$ (95\% CI: $[0.35,0.47]$), indicating a moderate positive association between retention stability and achievement growth.
Teacher persistence beyond year 3 is associated with an average increase of 9.3 points in annual student achievement growth (approximately 0.62 SD), accumulating to 31 points over four years (approximately 0.56 SD). These results support continuity and institutional knowledge mechanisms, whereby stable teacher retention enhances cumulative instruction and organizational capital. Figure~\ref{fig:studentMRL} illustrates student MRL recovery under the joint model.

%
%
%
\begin{table}[htbp]
	\centering
	\small
	\caption{Policy simulations: intervention timing, expected impacts, and cost-effectiveness.}
	\label{tab:policy_CE}
	\begin{tabularx}{\textwidth}{@{}>{\raggedright}p{4cm} c r r r@{}}
		\toprule
		Scenario & Timing & \makecell{Cost \\ (per teacher)} & \makecell{Tenure \\ gain} & \makecell{Net savings \\ (10 teachers / 8 yrs)} \\
		\midrule
		Front-loaded supports (mentoring + bonuses) & yrs 1--3 & \$15k & 1.8 yrs & \$127k \\
		Risk-targeted allocation ($<$50\% predicted 3-yr retention) & yrs 1--3 & \$8.5k & 2.3 yrs & \$183k \\
		Maintenance supports ($>$75\% predicted) & yrs 4+ & \$4.2k & 0.9 yrs & \$62k \\
		\bottomrule
	\end{tabularx}
\end{table}
\subsection{Policy Simulations and Cost-Effectiveness}
\label{sec:cost-effectiveness}

We conduct policy simulations that translate PMRL predictions and the empirically validated starshaped equilibrium into resource-targeting strategies. Guided by the burn-in period identified in Sections~\ref{sec:baseline-mrl} and \ref{subsec:empirical-test-starshaped}, we evaluate front-loaded interventions concentrated in years 1–3, consisting of mentoring, workload reductions, and targeted retention bonuses totaling \$15{,}000 per teacher. Using PMRL-based retention predictions, we further examine risk-stratified allocation, in which intensive supports target teachers with predicted three-year retention below 50\%, while maintenance supports are reserved for more stable cases.

Simulation results indicate that front-loaded supports yield positive returns by year 4 and generate cumulative net savings of approximately \$127{,}000 per 10-teacher cohort over an eight-year horizon through avoided replacement costs and continuity gains. Risk-targeted allocation improves cost-effectiveness further, producing larger tenure gains at lower per-teacher cost. Table~\ref{tab:policy_CE} summarizes intervention timing, expected tenure gains, and net fiscal benefits.
%
Overall, SMEL–PMRL outputs—particularly baseline MRL trajectories and post–burn-in expected tenure—map directly onto budgeting horizons and intervention design. Unlike hazard ratios, PMRL-based quantities support transparent fiscal planning, risk stratification, and sensitivity analysis. Results remain robust for replacement costs in the \$10{,}000–\$20{,}000 range and discount rates between 2\% and 5\%, with ROI remaining positive under empirically observed PMRL retention gains \parencite{CarverThomas2017,Ronfeldt2013}.

%


\section{ Discussion and Conclusions}
\label{sec:discussion} 

\subsection{Principal Contributions}

This study introduces the Starshaped Mean Residual Life (SMEL) framework to educational workforce research, adapting tools from reliability engineering to model teacher retention and student achievement as coupled dynamic equilibrium processes. Using extensive simulations and empirical analysis of data from 169 rural STEM teachers and 3{,}214 students, the proposed framework addresses long-standing methodological limitations in retention studies by shifting attention from short-term exit risks to the evolution of expected remaining career duration over time.

\paragraph{Theoretical Advancement.}
SMEL reconceptualizes teacher retention as an evolving equilibrium between investment forces—such as human capital accumulation, mentoring, and institutional support—and exit pressures including workload, isolation, and opportunity costs. The defining starshaped property, which requires the ratio $m(t)/t$ to be nondecreasing, provides a formal mathematical characterization of the transition from early-career vulnerability (``burn-in'') to a stable professional equilibrium. This equilibrium perspective moves beyond instantaneous hazard rates and instead focuses on expected remaining service, offering a more policy-relevant target for intervention design.

In contrast to Cox proportional hazards models, which rely on constant hazard ratios, and standard mean residual life models, which impose restrictive monotonicity assumptions, SMEL accommodates the nonmonotonic career trajectories observed in practice. These features are empirically justified by clear violations of proportional hazards in our data ($\chi^2_6 = 18.7$, $p = 0.005$). By directly modeling the mean residual life function $m(t)$, the framework yields quantities that are both statistically flexible and readily interpretable by policymakers and administrators.

\paragraph{Methodological Innovation.}
We extend the SMEL framework to regression settings through a proportional mean residual life (PMRL) specification,
$m(t \mid \mathbf{Z}) = m_0(t)\exp(\mathbf{Z}^\top \boldsymbol{\gamma}),$
estimated using a three-parameter Weibull--resilience baseline combined with adaptive Bayesian priors selected via Average Weighted Relative Efficiency (AWRE) criteria. Simulation studies demonstrate that SMEL--PMRL achieves a 5.4\% reduction in prediction error relative to Cox models, while maintaining negligible finite-sample bias (no larger than 0.02), even under heavy right-censoring of up to 40\%.
The framework further supports joint modeling of teacher retention and student achievement through a shared frailty structure, enabling direct quantification of their dependence. The estimated correlation between retention and achievement processes ($\hat{\theta} = 0.41$) highlights the interconnected nature of workforce stability and instructional outcomes. Complete and reproducible implementation code in \texttt{R} and \texttt{Stan} is provided to facilitate application and extension in other educational and workforce contexts.

\paragraph{Empirical Insights.}
Empirical application of SMEL--PMRL reveals substantial disparities in expected career duration for rural STEM teachers, who experience an average 32\% shorter remaining career length than their non-rural counterparts. The most pronounced losses occur during the first three years, when expected duration declines by 38\%, identifying a critical early-career intervention window. Policy simulations based on PMRL predictions show that front-loaded support packages generate positive returns on investment by year~4, aligning financial efficiency with workforce stabilization.

The analysis also uncovers meaningful heterogeneity in retention benefits. Participation in Noyce scholarship programs increases expected tenure by 47\%, while teachers with rural origins exhibit a 31\% advantage in career persistence. Importantly, retention gains extend beyond staffing outcomes: sustained teacher presence beyond year~3 is associated with annual student achievement gains of 9.3 points, accumulating to approximately 31 points over four years. These findings underscore the compounded educational and fiscal returns of early, targeted retention investments guided by the SMEL equilibrium framework.

\begin{table}[htbp]
	\centering
	\caption{Districts sorted by predicted retention $\hat{S}_d(3)$ from lowest to highest. Five moderate-risk 
		districts ($\hat{S}_d(3)  < 0.60$, top rows) require intensive early-career intervention during 
		years 1–3; stable/very stable districts ($\hat{S}_d(3) \ge 0.70$) require maintenance supports only.
		}
	\label{tab:district_summary}
	\begin{tabular}{lrrrrr}
		\toprule
		District & Teachers ($n_d$) & $\hat{S}_d(3)$ & Isolation (mean $Z_2$) & Noyce coverage (\%) & Risk band \\
		\midrule
		D4  & 21 & 0.475 &  2.06 & 23.8 & Moderate \\
		D3  & 22 & 0.475 &  1.79 &  4.5 & Moderate \\
		D2  & 14 & 0.480 &  1.80 & 14.3 & Moderate \\
		D7  & 22 & 0.575 &  1.09 & 22.7 & Moderate \\
		D5  & 24 & 0.575 &  1.03 & 20.8 & Moderate \\
		D6  & 10 & 0.607 &  1.10 & 30.0 & Stable \\
		D8  & 25 & 0.629 &  0.43 & 24.0 & Stable \\
		D12 & 24 & 0.708 &  0.02 & 66.7 & Stable \\
		D10 & 19 & 0.745 & -0.27 & 52.6 & Stable \\
		D11 & 11 & 0.762 & -0.59 & 36.4 & Very Stable \\
		D9  & 18 & 0.801 & -1.10 & 77.8 & Very Stable \\
		\midrule
		Overall & 210 & 0.635 &  0.60 & 33.8 & -- \\
		\bottomrule
	\end{tabular}
	\begin{tablenotes}
		\small
		\item \textit{Note:} $\hat{S}_d(3)$ computed by averaging PMRL-predicted teacher-level probabilities within districts. Correlation between $\hat{S}_d(3)$ and isolation: $r = -0.89$ ($p < 0.001$); correlation with Noyce coverage: $r = 0.85$ ($p < 0.001$).
	\end{tablenotes}
\end{table}
\subsection{Implications for Educational Policy and Practice}

\subsubsection{Strategic Resource Allocation}

SMEL-derived retention probabilities support evidence-based resource allocation across district, state, and federal levels by aligning intervention intensity with empirically identified risk.
At the district level, administrators can compute predicted retention trajectories using the fitted PMRL model to identify teachers and districts requiring front-loaded supports. Table~\ref{tab:district_summary} illustrates district-level variation in retention profiles, with predicted 3-year survival ranging from 0.475 to 0.801 across eleven participating districts. The five lowest-performing districts (D2--D7, $\hat{S}_d(3) < 0.60$) exhibit extreme geographic isolation (mean $Z_2 = 1.55$, corresponding to $>120$ miles from metropolitan centers) combined with minimal Noyce coverage (mean 17.2\%). These moderate-to-high-risk districts collectively employ 103 teachers and demonstrate retention probabilities 21--31\% below the sample mean, identifying them as priority targets for intensive early-career intervention during years 1--3.

District-level retention profiles are constructed by aggregating teacher covariates within districts and estimating predicted mean residual life $m(t \mid Z_{id})$ and three-year survival probabilities $\hat{S}_d(3)$. Districts with $\hat{S}_d(3) < 0.60$ are flagged for intensive supports (reduced teaching loads of 0.8 FTE, weekly mentoring, and quarterly retention bonuses totaling \$8,000--\$12,000 annually), while those exceeding 0.70 receive maintenance-level interventions. As summarized in Table~\ref{tab:district_summary}, predicted retention varies meaningfully across districts and aligns strongly with isolation indices ($r = -0.89$) and Noyce coverage ($r = 0.85$), reinforcing the value of SMEL-based risk stratification.

At the state level, SMEL predictions guide geographically targeted expansion of Noyce-style programs. Current funding reaches approximately 8\% of the rural STEM workforce; increasing coverage to 25\%—requiring an estimated \$45M annually in Texas—would stabilize roughly 140 additional teachers per year. Prioritizing expansion in the five moderate-risk districts identified in Table~\ref{tab:district_summary} would maximize return on investment, as these settings combine high attrition risk with demonstrated responsiveness to scholarship supports. SMEL-based cost-benefit analyses indicate a 3.2:1 return through avoided replacement costs and student achievement gains.
At the federal level, SMEL benchmarks provide an alternative to uniform retention targets. Rather than imposing fixed thresholds (e.g., 80\% retention at five years), NSF program officers can evaluate grantee performance relative to context-adjusted expectations accounting for geographic isolation, labor market conditions, and student demographics, preserving accountability without penalizing programs in high-need settings. Table~\ref{tab:district_summary} demonstrates this need: Districts D3 and D9 differ by 33 percentage points in predicted retention (0.475 vs.\ 0.801), yet contextual factors—isolation differential of 2.89 standard deviations and 73-point Noyce coverage gap—fully explain this disparity, suggesting both programs operate effectively within their constraints.
\subsubsection{Intervention Design Principles}

The SMEL framework yields four empirically grounded principles for designing retention programs that align intervention intensity with observed risk patterns.

First, intervention intensity should be front-loaded during the early-career 
burn-in phase (years 1–3), when expected remaining tenure declines most rapidly. 
SMEL-aligned interventions for moderate-risk districts ($\hat{S}_d(3) < 0.60$) include 
reduced teaching loads (0.8 FTE), weekly mentoring, monthly peer learning 
communities, and quarterly retention bonuses totaling \$8,000–\$12,000 annually.

Second, SMEL predictions enable risk-differentiated allocation at both teacher and district levels. Districts with $\hat{S}_d(3) < 0.60$ require comprehensive, front-loaded supports, while districts with $\hat{S}_d(3) \ge 0.70$---classified as stable or very stable in Table~9---require maintenance-level interventions only. This stratification improves allocative efficiency under fixed budgets by concentrating resources where marginal returns are highest. In the current sample, the five moderate-risk districts collectively employ 103 teachers and exhibit retention probabilities 21--31\% below the sample mean, justifying priority targeting during years~1--3.

Third, the strong effect of geographic isolation ($\exp(\gamma)=0.81$ per standard deviation) indicates that individual-level supports alone are insufficient in high-risk settings. Moderate-risk districts identified in Table~9 exhibit extreme isolation (mean $Z_2 = 1.55$), corresponding to distances exceeding 120 miles from metropolitan centers. Effective intervention portfolios must therefore address structural barriers through spousal employment assistance, housing subsidies that reduce effective isolation, technology-enabled remote mentoring, and regional professional learning communities that mitigate geographic constraints.

Finally, the 31\% tenure advantage observed among rural-origin teachers ($\exp(\gamma)=1.31$) underscores the efficiency of ``grow-your-own'' recruitment strategies as a complementary long-term solution. District investments in local scholarship pipelines---approximately \$60{,}000 per teacher over four years---yield retention rates two to three times higher than external recruitment. Coordinated state-level expansion of such pipelines, guided by SMEL-based district risk profiles, can amplify these gains while limiting inefficient competition among neighboring districts.

\subsubsection{Beyond Rural STEM: Transferable Framework}

Although validated in rural STEM education, the SMEL-PMRL framework generalizes to other high-attrition workforce settings. School leadership turnover exhibits similar burn-in dynamics and could benefit from SMEL-informed succession planning. Nursing faculty shortages parallel STEM teacher attrition, with SMEL offering tools to evaluate workload and salary differentials. Visa-dependent international teachers face distinct retention barriers that SMEL can quantify across policy regimes. More broadly, nonprofit organizations may use SMEL to distinguish mission-complete exits from burnout-driven attrition, enabling more targeted retention strategies.

\subsection{Limitations and Boundary Conditions}

Our study has several limitations that warrant consideration. First, \text{generalizability is constrained} by the geographic and programmatic context. The empirical analysis draws on rural Texas districts from 2018--2023, a period including the COVID-19 pandemic. 
Labor market conditions, state policies, and the specific implementation of the NSF 
Noyce scholarship program may differ elsewhere, limiting direct extrapolation to 
other rural regions or time periods. The 2018–2023 observation window includes the 
COVID-19 pandemic (2020–2021), which may have temporarily altered retention dynamics 
through emergency credentialing, remote instruction adaptations, or heightened 
occupational stress; however, post-pandemic patterns (2022–2023) showed convergence 
toward pre-2020 trajectories. Multi-state, long-term replication is needed.
Second, \text{measurement limitations} affect the interpretation of key constructs. The primary retention outcome treats all departures as equivalent, obscuring differences between voluntary exits, dismissals, and within-profession transfers. Student achievement is measured via standardized tests, capturing only a narrow dimension of STEM learning. Most notably, the geographic isolation index—a standardized distance-to-metro measure—serves as a proxy that likely conflates physical remoteness with correlated structural disadvantages (e.g., spousal employment opportunities, school resources, poverty). The estimated 19\% reduction in expected tenure per standard deviation of isolation may thus reflect a composite of barriers, 
complicating causal interpretation and precise policy targeting. Third, the modeling 
framework relies on several statistical assumptions. The SMEL-PMRL model assumes 
independent censoring and a proportional effect of covariates on the mean residual 
life across all career stages. The joint teacher–student model assumes normally 
distributed shared frailties. While sensitivity analyses support the robustness of 
our conclusions, violations of these assumptions (e.g., time-varying covariate 
effects, informative censoring, or skewed frailty distributions) could affect the 
results. Future methodological extensions could relax these assumptions.
\subsection{Concluding Synthesis}

Persistent shortages of qualified STEM teachers in rural districts remain a critical barrier to educational equity in the United States. Despite sustained investments through programs such as the NSF Robert Noyce Teacher Scholarship Program, the literature has lacked quantitative models capable of capturing the nonlinear attrition patterns characteristic of high-need rural contexts. 
This study addresses that gap by adapting the Starshaped Mean Residual Life (SMEL) framework from reliability engineering to educational workforce dynamics.
SMEL reconceptualizes teacher retention as a dynamic equilibrium between investment forces and exit pressures, formally capturing the transition from early-career ``burn-in'' vulnerability to long-run stability. The starshaped condition—requiring the ratio $m(t)/t$ to be nondecreasing—provides a rigorous equilibrium criterion and overcomes the restrictive assumptions of proportional hazards models that obscure heterogeneous career trajectories.
Methodologically, the proportional mean residual life (PMRL) regression framework, combined with adaptive Bayesian estimation and a three-parameter Weibull-resilience specification, enables flexible modeling of non-monotonic hazard structures while preserving interpretability. Simulation results demonstrate that SMEL-PMRL reduces prediction error by 5.4\% relative to Cox models and maintains minimal bias ($\leq 0.02$) under right-censoring up to 40\%. Extensions to joint teacher--student models using shared frailty quantify feedback between workforce stability and instructional effectiveness.
Empirical analysis of 169 rural STEM teachers and 3,214 students shows that rural teachers experience 32\% shorter expected career duration than non-rural peers, with the steepest decline (38\%) concentrated in years 1--3. This pattern identifies a narrow but critical intervention window: intensive early-career supports maximize returns, while maintenance-level interventions suffice once teachers reach post-equilibrium stages (years 4+). Persistence beyond year 3 yields substantial student achievement gains—9.3 points annually (0.62 SD), accumulating to 31 points (0.56 SD) over four years.

These findings translate directly into policy guidance. SMEL-derived retention probabilities enable evidence-based targeting of supports, prioritizing early-career teachers at highest risk while avoiding inefficient uniform allocation. Cost-effectiveness simulations indicate that concentrated supports totaling \$15{,}000 per teacher during years 1--3 achieve positive ROI by year 4 and generate approximately \$127{,}000 in net savings per 10-teacher cohort over eight years. Structural interventions addressing geographic isolation, spousal employment constraints, and local recruitment pipelines further amplify these gains.

Beyond rural STEM education, SMEL-PMRL offers a transferable framework for workforce retention in other high-attrition settings, including school leadership, nursing education, and nonprofit organizations. By centering inference on expected remaining service rather than instantaneous risk, SMEL translates technical output into policy-relevant quantities that directly inform hiring, budgeting, and program design.
Limitations include geographic specificity to rural Texas, focus on NSF Noyce participants, and a five-year observation window. Future research should examine multi-state cohorts, longer follow-up horizons, and experimental interventions, along with methodological extensions incorporating time-varying covariates, spatial dependence, and causal mediation.
Overall, this study establishes SMEL-PMRL as a statistically rigorous, empirically validated, and policy-relevant framework for analyzing educational workforce dynamics. By bridging reliability theory and education policy analysis, it provides actionable tools for identifying intervention windows, allocating resources efficiently, and strengthening STEM teacher retention in high-need rural systems.

\section*{Conflicts of Interest}
The author declares no conflicts of interest related to this research. 
 All participants provided informed consent. 
De-identified data and analysis code will be available from the corresponding author upon request.

\printbibliography

\end{document}